\documentstyle[12pt,psfig]{article}
\begin{document}
\baselineskip 6mm
{\noindent  \large \bf
{MANY-ELECTRON WAVE FUNCTION AND MOMENTUM \\ DENSITY}
\vspace{6mm}}

\noindent{\large
B. Barbiellini}
\vspace{2mm}

{\small
\noindent
Department of Physics, Northeastern University, Boston MA 02115
\vspace{2mm}
\vspace{6mm}

\noindent
Inelastic x-ray scattering at large momentum transfer is an ideal probe
of the ground state of 
electrons in condensed matter. The experimental determination of the electron momentum 
density (EMD) is based on the Impulse Approximation (IA).  In general, the EMD even in 
simple metals cannot be well represented by the mean-field Independent Particle Model 
(IPM). In other words, the many-electron wave function is not well described by a single 
Slater determinant.  Instead the momentum density has to be constructed from a correlated 
state with average occupancies in between 0 and 1.  For the Homogeneous Electron Gas 
(HEG) considerable effort has been made to deduce the size of the discontinuity $Z$ at the 
Fermi momentum. The difference between the interacting and free HEG momentum 
distributions for several electron densities yields the Lam-Platzman Correction (LPC) within 
the Density Functional Theory (DFT). Since the Quantum Monte Carlo (QMC) method 
applied to the HEG is used to determine DFT correlation  potentials \cite{ortiz}, 
a consistent treatment 
of the LPC should utilize the results of these same QMC simulations. The LPC gives a 
redistribution of the EMD from small momentum regions to larger momentum values. It 
describes some interactions between the electrons beyond the IPM. However, the LPC is 
spherically symmetric while the correction in real materials should be anisotropic. For more 
realistic corrections, one can perform QMC or GW calculations. These calculations are much 
more time consuming than the DFT computations. Moreover it was found that QMC 
Compton profiles are still to high at zero momentum as compared to experiments in materials 
ranging from lithium \cite{filippi}
to silicon \cite{louie}. A recent GW calculation for lithium \cite{eguiluz} agrees with 
the QMC simulation.  This indicates that both the QMC and the GW methods do not capture 
all the effects observed with the Compton scattering.  These effects could include, in some 
extent, the failure of the IA or multiple scattering corrections. However, Bouchard and 
Lhuillier have shown that the momentum density is also extremely sensitive to the way the 
anti-symmetry is implemented in the many-fermion wave function \cite{bouchard}. The current QMC and 
GW calculations assume that fermionic correlation is not catastrophically modified by 
interactions and that an adiabatic path exists between the free electron gas 
and the interacting 
liquid. In the QMC method, this assumption leads to the fixed node approximation. The 
standard picture of the interacting gas appears to be substantially correct in the aluminum 
Compton profile \cite{suortti}, but in other materials some deviations from may occur as soon as the 
Fermi surface is not rotationally symmetric. Two recent examples are given by the 
experimental Compton profiles of beryllium \cite{huotari} and copper \cite{sakurai}.  
It is therefore worthwhile to 
investigate other ways to implement anti-symmetry such as the Antisymmetrized Geminal 
Product (AGP) \cite{agp}. The AGP yields an orbital-dependent approach in which the momentum 
density is constructed using the natural orbitals, and the corresponding occupation numbers 
are obtained through a variational procedure. Sometime it is advantageous to introduce the 
Wannier orbitals via the natural orbitals \cite{goedecker} in order to study the degree of locality of 
bonding properties. The range of the Wannier orbitals is a fundamental ingredient for 
understanding the existence of insulators and metals and, in particular, the Metal-Insulator 
Transition (MIT). Electron localization in a MIT can be produced both by correlation effects 
and by disorder in an otherwise ideal crystal lattice. Actually, the path integral formalism  
\cite{fulde}
connects the theory of interacting electrons to that of disordered systems, 
where electrons are 
moving in randomly distributed external potentials. Thus, correlation and disorder can 
produce some similar effects in the EMD.
This work is supported in part by the US Department of Energy contract
W-31-109-ENG-38, and the allocation of supercomputer time at the 
Northeastern University Advanced Scientific Computation Center (NU-ASCC).


\begin{thebibliography}{99}
\bibitem{ortiz}
G. Ortiz and P. Ballone, Phys. Rev. B 50 (1994) 1391.\\
\bibitem{filippi}
C. Filippi and D.M. Ceperley,  Phys. Rev. B 59 (1999) 7907.\\
\bibitem{louie}
B. Kralik, P. Delaney and S. G. Louie,  Phys. Rev. Lett. , 80 (1998) 4253.\\
\bibitem{eguiluz}
A.G. Eguiluz, W. Ku, J.M. Sullivan, J. Phys. Chem. Solids 61 (2000) 383.\\ 
\bibitem{bouchard}
J.B. Bouchaud and C. Lhuillier, Z. Phys. B 75 (1989) 283.\\
\bibitem{suortti}
P. Suortti et al., J. Phys. Chem. Solids 61 (2000) 397.\\
\bibitem{huotari}
S. Huotari et al, preprint.\\
\bibitem{sakurai}
Y. Sakurai et al., J. Phys. Chem. Solids 60 (1999) 905.\\
\bibitem{agp}
B. Barbiellini, J. Phys. Chem. Solids 61 (2000) 341.\\
\bibitem{goedecker}
S. Goedecker and E. Koch, preprint cond-mat/9905208.\\
\bibitem{fulde}
P. Fulde, Electron Correlations in Molecules and Solids, 3, 
Springer, Berlin, 1995.\\
\end{thebibliography}
\end{document}